\documentclass[aps,prl,superscriptaddress,floatfix,nofootinbib,showpacs,longbibliography]{revtex4-2}

\usepackage{graphicx}
\usepackage{amsmath,amsthm,amssymb}
\usepackage{color}
\usepackage{multirow}
\usepackage[utf8]{inputenc}
\usepackage{tabularx}
\usepackage{booktabs}
\usepackage{longtable, tabu}
\usepackage{ltablex}
\usepackage{ltxtable}
\makeatletter
\def\be{\begin{equation}}
	\def\ee{\end{equation}}
\def\bea{\begin{eqnarray}}
	\def\eea{\end{eqnarray}}
\def\ben{\begin{equation*}}
	\def\een{\end{equation*}}
\def\bean{\begin{eqnarray*}}
	\def\eean{\end{eqnarray*}}
\def\bma{\begin{mathletters}}
	\def\ema{\end{mathletters}}
\def\bi{\begin{itemize}}
	\def\ei{\end{itemize}}
\def\bnu{\begin{enumerate}}
	\def\enu{\end{enumerate}}

\newcommand{\ket}[1]{ | \, #1  \rangle}
\newcommand{\bra}[1]{ \langle #1 \,  |}
\newcommand{\proj}[1]{\ket{#1}\bra{#1}}

\begin{document}

\title{Asymptotically secure All-or-nothing Quantum Oblivious Transfer}

\author{Ramij Rahaman}
\email{ramijrahaman@isical.ac.in} \affiliation{Physics \& Applied Mathematics Unit, Indian Statistical Institute, 203 B. T. Road, Kolkata 700108, India}
	
\begin{abstract}
	We present a device independently secure quantum scheme for $p$-threshold all-or-nothing oblivious transfer. Novelty of the scheme is that, its security does not depend - unlike the usual case - on any quantum bit commitment protocol, rather it depends on Hardy's argument for two-qubit system. This scheme is shown to be unconditionally secure against any strategy allowed by quantum mechanics. By providing a secure scheme for all-or-nothing quantum oblivious transfer, we have answered a long standing open problem, other than the quantum key distribution, whether there is any two-party quantum cryptographic protocol, which is unconditionally secure. 
	
\end{abstract}


\maketitle	

\section{Introduction}

If Alice and Bob have access to a preshared key for authentication, they can establish a shared secret key by using an insecure quantum channel and public communication - a scheme called as quantum key distribution (QKD), and the unconditional security of QKD is now well accepted \cite{BB84,ABGSPS07,VV14,RPMP15}. Barring the grand success of QKD there are several important two-party computational cryptographic primitives like bit commitment, oblivious transfer, {\em etc.}, where the role of quantum mechanics is still not clear \cite{May97,LC97,Lo97,WTHR11,BCS12}. In fact, in some cases, quantum mechanics has already provided the no-go proofs for them. Quantum bit commitment(QBC) is one of such protocol \cite{May97,LC97}, and it was shown that an unconditionally secure bit commitment protocol is impossible irrespective of whether one only uses classical or quantum communication; hence several proposed quantum oblivious transfer (QOT) which, if based on QBC, cannot be secure either \cite{Lo97,He12}. Thus, an important question remains unanswered: Is there any two-party cryptographic protocol other than QKD which can be unconditionally secure? Our answer is in the affirmative, and to establish our claim, here we provide a perfectly secure QOT scheme which is not based on QBC.

The concept of original oblivious transfer (OT), first introduced by Rabin \cite{Rab81}, is: Alice sends an one bit of message $(m)$ to Bob and she knows that with probability $\dfrac{1}{2}$ Bob can decrypt the bit $m$, but this is the only thing she would ever know. Whereas Bob knows when his decryption is successful, he is 100\% sure, but his success probability is $\dfrac{1}{2}$ and Alice is totally ignorant about his success. If we replace the probability $\dfrac{1}{2}$ with the probability $p$ that Bob successfully decrypts the message $(m)$ which he received from Alice, then this OT protocol is generally called a $p$-OT. Another type of OT called {\em 1-out-of-2} Oblivious transfer $\left( {1\choose 2}-\text{OT}\right)$, was invented by Even, Goldreich and Lempel \cite{EGL83}. In this scenario, Alice allows to transfer to Bob exactly one bit of secret, out of two recognizable secret bits $b_0$ and $b_1$ and Bob can decrypt bit $b_k$ and not $b_{\overline{k}}$ with probability $\dfrac{1}{2}$. Bob knows (with certainty) which of $b_0$ or $b_1$ he got, whereas Alice does not know which $b_0$ or $b_1$ Bob has got. In 1997, Lo \cite{Lo97} provided a no-go security proof for a broad class of two party quantum protocols. An immediate corollary to this result is the impossibility of an unconditionally secure quantum $ {1\choose 2}$-OT as it would imply a secure scheme for QBC. In 1987, Cr\'{e}peau \cite{Cre87} showed that Rabin's $p$-OT and $ {1\choose 2}$-OT are equivalent for all values of $p (0<p<1)$ - each one of them can be implemented by using the other as a primitive. Latter, He and Wang showed that the two `flavors' of oblivious transfer protocols, namely, $p$-OT and ${1\choose 2}$-OT, are no longer equivalent in the quantum domain unlike its classical case \cite{HW06n,PPP17}. Thus, a secure quantum $p$-OT is possible \cite{HW06} and have no contradiction with the existing no-go proof of secure QBT protocol.

Many attempts were made by several researchers to construct a unconditionally secure QOT protocols but none of them got success till date \cite{BBCS92,Yao95,BCW03}. But, several recent results show that secure quantum protocols for $\frac{1}{2}$-OT and $ {1\choose 2}$-OT are possible \cite{SMAP15,He15,SMAP15r,PPP17} with negligible imperfection \cite{Ami21,KST20}. In recent time QOT schemes have also been studied under space-time paradigm \cite{Pit16,PK18,Pit19}. In this regard, we present a device independently secure QOT, the secrecy of which is based on the unique features Hardy's paradox \cite{Har92} for two-qubit system. We mainly discuss here about the implementation of Rabin's $p$-OT protocol as, from theoretical point of view, there is no difference with the $\frac{1}{2}$-OT, rather it's a more general one.

 In 2013, Chailloux {\em et al.} \cite{CKS13} claimed that a perfectly secure $ {1\choose 2}$-OT quantum protocol is impossible. They derived a lower bound for QOT based on some earlier works on {\em coin flipping} \cite{Kit03} and {\em bit commitment} \cite{CK11}. Our secure QOT protocol does not invalidate none of those works rather the generality of their assumptions. The class of quantum
protocols considered in these works was too narrow and all impossibility proofs were relying on non-uniqueness of states in the
considered classes of protocols. Whereas in our case, Hardy paradox uniquely defines the state and both the party can check the genuineness of the Hardy state by simply exchanging classical information.

We organize our paper as follows: we start with a description of special quantum correlations, leading to the so-called Hardy paradox \cite{Har92}, which will allow us to formulate the QOT protocol. Next we present the protocol and discuss several security aspects along with device independent (DI) scenario and its noise robustness, and finally we end with a conclusion.

\section{Hardy's paradox}
Consider a bipartite physical system consisting of
two subsystems each one being possessed separately by two distant parties Alice and Bob. Assume that
Alice can run the experiments of measuring one (chosen at
random) of the two $\{-1,+1\}$-valued random variables $U_1$ and $D_1$
on the states of her subsystem. Similarly, Bob can also run the experiments
of measuring any one (chosen at random) of the two $\{-1,+1\}$-valued
random variables $U_2$ and $D_2$ on the states of his subsystem. The Hardy-type argument consists of the following set of four joint probability
conditions for the bipartite system:
\be\label{hardy2q}
\begin{split}
P(+1,+1|U_1,U_2) &= q>0,\\
P(+1,+1|U_1,D_2) &= 0,\\
P(+1,+1|D_1,U_2) &= 0,\\
P(-1,-1|D_1,D_2) &= 0.	
\end{split}
\ee
This set of conditions cannot be satisfied by any {\em local-realistic} (LR) theory but can be satisfied in quantum theory \cite{Har92}. But, an unique two qubit non-maximally entangled state known as Hardy state $\ket{\psi^H}$ satisfies (\ref{hardy2q}) for a given set of local observables pairs $\{(U_1,D_1),(U_2,D_2)\}$ \cite{Kar97}. In a two-qubit system the maximum probability
of success of Hardy's argument \ref{hardy2q} is $q=q_{max}= \dfrac{5\sqrt{5}-11}{2}\approx 0.09$, which occurs when $U_1\equiv U_2$ ($=U$, say) and $D_1\equiv D_2$ ($=D$, say) \cite{Jor94}. Let $\ket{x}$ ($\ket{x^\perp}$) denotes the eigenstate of the observable $X\in \{U,D\}$ with eigenvalue $+1(-1)$. Now expressing $D$ in terms of the orthogonal basis of $U$ i.e.,
\be\label{obser2}
\begin{split}
	\ket{d}&=\alpha\ket{u}+\beta \ket{u^\perp},\\ \mbox{~and,~}\ket{d^\perp}&=\beta^*\ket{u}-\alpha^* \ket{u^\perp},
\end{split} \ee
where $\displaystyle |\alpha|^2+|\beta|^2=1$ and $0<|\alpha|<1$, one can easily observe that 

	\be\label{hardy_state}
\ket{\psi^H}=c_{1}\ket{d}\ket{u^\perp}+c_{2}\ket{d^\perp}\ket{u}+c_{3}\ket{d^\perp}\ket{u^\perp},
\ee
where, $c_{1}=\frac{-{\alpha^*}^2\beta}{|\alpha|^2\sqrt{1-|\alpha|^4}}$, $c_{2}=\frac{\beta|\alpha|^2}{\sqrt{1-|\alpha|^4}}$, $c_{3}=\frac{\alpha^*|\beta|^2}{\sqrt{1-|\alpha|^4}}$, and consequently, $\displaystyle q=\dfrac{|\alpha|^4|\beta|^2}{1+|\alpha|^2}$ \cite{RWZ15}.

Ref. \cite{RZS12} tells that the maximum probability of success  $q_{max}=\frac{5\sqrt{5}-11}{2}$ leads to a DI test for the state $\ket{\psi^H}$.

\section{Quantum Oblivious transfer(QOT)}
Let us first describe the quantum protocol for $p$-OT in details. Rabin's original OT protocol is just a particular case of $p$-OT with $p=\dfrac{1}{2}$, and it can always be implemented from a $p$-OT for $0<p<1$ \cite{Cre87}. Let two distant parties Alice and Bob want to implement a secure $p$-QOT scheme for them and they have been provided with the facility of public communication as well as noiseless transportation of a physical system like a spin-$1/2$ system or a polarized photon. The $p$-QOT protocol for them works as follows:   

\begin{center}
		
	\begin{table}[h]
		\caption{Outline of the quantum $p$-OT protocol}
		\label{pOT}
		\begin{center}
			\begin{tabular}{  c  p{\textwidth} }
				\toprule
				\textbf{Steps}      
				& ~~~~~~~~~~~~~~~~~~~~~\textbf{Resource preparation}   \\
				\midrule
				
				S1.& Alice prepares a large number $N$ of copies of Hardy state $\ket{\psi^H}$ and shares with Bob. \\
				S2. & Bob randomly chooses whether to measure $U$, or
				$D$ on each of his qubits. Alice does the same and right after measurement she sends her qubit to Bob. In each run, both Alice and Bob keep their individual record of the chosen measurements $A_i$ and $B_i$ and the corresponding measurement outcomes $a_i$ and $b_i$, respectively.\\
				S3.& Bob declares the list $L^+$ contains all the runs for which his measurement result was $b_i=+1$.  \\
				S4. & Alice prepares (randomly) a list $\{R_i\}$ of pairs of runs from $L^+$, where for each pair of runs $R_i=(i_1,i_2)$ with $A_{i_1}=A_{i_2}$ and $a_{i_1}\neq a_{i_2}$. Alice sends the list $\{R_i\}$ of pairs of runs to Bob.  \\
				S5. & Bob announces all the pairs of runs (say, $\{R'_i\}\subset \{R_i\}$), in which his measurement pairs are different {\em i.e.} $B_{i_1}\neq B_{i_2}$.\\
				
				\midrule				     
				& ~~~~~~~~~~~~~~~~~~~~~\textbf{OT encoding and decoding}   \\
				\midrule
				S6.& Alice randomly selects one pair of runs from $\{R'_i\}$, say $R_{i^*}=(i^*_1,i^*_2)$, according to the message she wants to convey to Bob (say, ``Alice chooses $(A_{i^*_1},A_{i^*_2})=(U,U)$ for $0$, while she chooses $(A_{i^*_1},A_{i^*_2})=(D,D)$ for $1$"), and announces only the index $i^*$ for OT. \\
				
				S7. & Bob performs measurements in $X_{j}=B_{i^*_j}$ bases 
				on the $i^*_j$-th qubit (for $j=1,2$), he received from Alice in step S2. \\
								\bottomrule
			\end{tabular}
			
		\end{center}
	\end{table}	
	
\end{center}

\bnu
	\item[ S1.] {\em Source, qubit distribution:}
	Alice prepares a large number ($N$, say) of two qubit non-maximally entangled {\em Hardy states} $\ket{\psi^H}$ given in Eq. (\ref{hardy_state}), which satisfies all the conditions of (\ref{hardy2q}) for the observables pairs $(U,D)$ (on each local site). From pair of qubits ($i_A$ and $i_B$) for each $\ket{\psi^H}_i$ ($i=1,2,\dots,N$), Alice sends one qubit ($i_B$) to Bob, and
	keeps the other qubit ($i_A$) for herself.
	
	\item[S2.] {\em Local measurements and sending the measured qubits:}
	Bob randomly chooses whether to measure $U$, or $D$ on his qubit $i_B$ associated with the $i$-th {\em Hardy state}. Alice does the same: chooses randomly between measurements of $U$ and $D$ and right after measurement, she sends her measured qubit to Bob and this has to be done within the individual time window. The entire process of (i) sending one of the qubits of the two-qubit state $\ket{\psi^H}$ to Bob, (ii) performing the individual measurements on their respective qubits by Alice and Bob, and (iii) there by sending the qubit to Bob, should constitute one time window. There is no overlap among these time windows. So each time window can be referred to as a {\em run}. In the case of each run, $i$, they write down not only the chosen observables, $A_i$ and $B_i$ respectively, but also the obtained results, $a_i$ and $b_i$. Alice may be allowed to send a different qubit at each run whose state is same as that of the original qubit after her measurement but not any other qubits or a qubit which is already correlated with other system. At a later verification time Bob can check whether he received the actually measured qubits or it's replica (as mentioned above) for each run, or not.
	
	\item[S2(a).] {\em Honesty check of Alice's source, and of Alice's and Bob's actions:} Alice randomly selects some runs $R_A$ ({\em i.e.}, for some randomly selected values of $i$), and asks Bob to announce his measurement choices $B_i$ and the corresponding outcomes $b_i$ for those runs $R_A$. Similarly Bob also randomly selects some runs $R_B$, and asks Alice to announce her measurement choices $A_i$ and the corresponding outcomes $a_i$ for those runs $R_B$. 
	A state other than $\ket{\psi^H}$ cannot satisfy all Hardy conditions in (\ref{hardy2q}) for the given pairs of local observables $(U,D)$ on each side \cite{FtNote}. All possible interventions of Eavesdropper, which may include cheating by Alice or Bob, {\em e.g.} by establishing any type of correlations by coupling to the state $\ket{\psi^H}$, or emitting a different state, or using different measurement settings other than the ones of the protocol, can be found out. Alice and Bob simply publicly compare their announced measurement choices and the corresponding outcomes with the Hardy conditions (\ref{hardy2q}). Thus, in this case, assuming the ideal situation, {\em i.e.}, Alice and Bob to be honest and assuming no {\em Eavesdropping}, the possible two-qubit states with Bob (after S2) in each run $i$ will be one of
	\bean
	&\ket{u}\ket{u},\ket{u}\ket{u^\perp},\ket{u}\ket{d^\perp},\ket{u^\perp}\ket{u},\ket{u^\perp}\ket{u^\perp},\ket{u^\perp}\ket{d},\ket{u^\perp}\ket{d^\perp},&\\
	&\ket{d}\ket{u^\perp},\ket{d}\ket{d},\ket{d}\ket{d^\perp},\ket{d^\perp}\ket{u},\ket{d^\perp}\ket{u^\perp},\ket{d^\perp}\ket{d}.&
	\eean
	Bob also can check whether he received Alice's actual qubits at step S2, by performing the same measurements as those of Alice on qubits he received from Alice for those announced runs and compare the outcomes with Alice's outcomes. If not all the qubits he received from Alice are genuine then Bob cannot have consistent outcomes with the outcomes announced by Alice for those runs. Sending entangled qubits, which are correlated with ancilla systems at Alice's place is even more useless, as Bob would get results of measurements on such qubits inconsistent with Hardy's conditions. Thus, the above check fully
	ensures that qubits, which Bob received from Alice are all genuine.
	
	\item[S3.] {\em Runs offered by Bob:} Define the set of runs ``$L=\{\mbox{All runs}\}\smallsetminus \{\mbox{Announced runs in S2(a)}\}$". From list $L$, Bob prepares a new list $L^+$, such that $L^+$ contains all the runs of $L$ for which his measurement outcome was $b_i=+1$. Thus, in this case, assuming the ideal situation, the possible two-qubit states with Bob now (in each run $i$ from the list $L^+$) will be one of
	\be\label{S4}
	\begin{split}
		\ket{u}\ket{u},\ket{u^\perp}\ket{u^\perp},\ket{u^\perp}\ket{d},\\
		\ket{d}\ket{d},
	\ket{d^\perp}\ket{u},\ket{d^\perp}\ket{d}.	
\end{split}
	\ee
	Bob now announces the values of $i$ for the list $L^+$, without revealing the observables $B_i$. Examples are given in Table \ref{TableS3}.
\begin{table}[h]
	\caption{Table for runs $L^+$ (step S3)}\label{TableS3}
	\begin{tabular}{ |c|c|c||c|c|c|c| }
		\cline{1-7}
		{Run} & \multicolumn{2}{|c||}{Alice (at S3)} & \multicolumn{2}{c|}{Bob (at S3)} & List & List  \\
		\cline{2-5}
		$(L^+)$  & Measurement  & Outcome & Measurement & Outcome &  $\{R_i\}$  & $\{R'_i\}$\\
		& basis$(A_i)$ & $(a_i)$ & basis$(B_i)$ & $(b_i)$ & (S4) & (S5) \\
		\cline{1-7}
		\vdots      & \vdots   &\vdots     & \vdots     &   \vdots    &     &        \\
		\cline{1-7}
		$i_1$    & U   &$-1$    & D & $+1$  & $\in R_1$     & $\in R'_1$\\
		\cline{1-7}
		$i_2$    & U   &$+1$    & U & $+1$  & $\in R_1$     & $\in R'_1$\\
		\cline{1-7}
		$i_3$    & D   &$-1$    & D & $+1$  & $\in R_2$     &\\
		\cline{1-7}
		$i_4$    & U   &$-1$    & D & $+1$  & $\in R_3$     & $\in R'_2$\\
		\cline{1-7}
		$i_5$    & D   &$+1$    & D & $+1$  & $\in R_2$     &\\
		\cline{1-7}
		$i_6$    & U   &$+1$    & U & $+1$  & $\in R_3$     & $\in R'_2$\\
		\cline{1-7}
		$i_7$    & U   & $-1$   & U & $+1$  & $\in R_4$     &\\
		\cline{1-7}
		$i_8$    & D   &$+1$    & D & $+1$  & $\in R_5$     & $\in R'_3$\\
		\cline{1-7}
		$i_9$    & D   & $-1$   & U & $+1$  & $\in R_5$     & $\in R'_3$\\
		\cline{1-7}
		$i_{10}$& U   & +1     & U &$+1$   & $\in R_4$     &\\
		\cline{1-7}
		\vdots      & \vdots   & \vdots     & \vdots&  \vdots     &              &\\
		\cline{1-7}
	\end{tabular}
\end{table}
	
	\item[S3(a).] {\em Alice's check of Bob's honesty in S3:} Bob might have included some runs in $L^+$, where his outcomes are $b_i=-1$. If he did that in S3 then that can be found out by Alice's check. Alice simply asks Bob to reveal his measurement bases for some randomly selected runs from the list $L^+$, say $25\%$ (the selected set will be called $L^+_1$). By comparing with her own measurement results, for runs $i$ in $L^+_1$, Alice can test whether the list $L^+$ are genuine list of measurements of protocol observables performed on the {\em Hardy state}. If this is the case, due to the second and the third condition of (\ref{hardy2q}), for the list $L_1^+$, Alice would never find `$U=+1$' (or, `$D=+1$') if Bob's result is `$D=+1$' (or, `$U=+1$'). Whereas, for the {\em Hardy state} associated to the observables pair $(U,D)$ on each side, $P(-1,+1|D,U)=\frac{|\alpha|^4}{1+|\alpha|^2}>0$ and $P(-1,+1|U,D)=\frac{1}{1+|\alpha|^2}>0$.
	
	As the qubits are in Bob's hand, he may also cheat by preparing the list $L^+$ in such a way that it contains only those runs where his two-qubit measurement outcomes on states of (\ref{S4}) are either $\ket{u^\perp}  \ket{d}$ or $\ket{d^\perp}  \ket{u}$. But this will cause a list with very few runs compared to the actual list $L^+$. Note that $P_B(+)=\frac{1+|\alpha|^4}{2(1+|\alpha|^2)}$ (the probability that Bob's outcome is $+1$) whereas in cheating case the probability is strictly less than $\frac{1}{2}P_B(+)$. Alice can check, whether the list $L^+$ contains enough runs ({\em i.e.}, $P_B(+)$-times of the total runs of $L$) or not to match the desire probability.
	
	\item[S4.] {\em Runs offered by Alice:}
	Define the set of runs ``$L^+_2=\{L^+\smallsetminus L^+_1\}$". From $L^+_2$, Alice prepares (randomly) a list of pairs of runs, where for each pair of runs $R_i=(i_1,i_2)$ with $A_{i_1}=A_{i_2}$ and $a_{i_1}\neq a_{i_2}$. Thus for each pair $R_i=(i_1,i_2)$ from $L^+$, the possible pairs of joint states of Alice and Bob will be
	\be\label{forOT}
	\begin{split}
		\left(\ket{u}\ket{u},\ket{u^\perp}\ket{u}\right),\left(\ket{u^\perp}\ket{u},\ket{u}\ket{u}\right),\\
		\left(\ket{d}\ket{d},\ket{d^\perp}\ket{d}\right),\left(\ket{d^\perp}\ket{d},\ket{d}\ket{d}\right),\\
	(\ket{u}\ket{u},\ket{u^\perp}\ket{d}),(\ket{u^\perp}\ket{d},\ket{u}\ket{u}),\\(\ket{d}\ket{d},\ket{d^\perp}\ket{u}),\left(\ket{d^\perp}\ket{u},\ket{d}\ket{d}\right).
\end{split}
	\ee
	Alice now announces the list $\{R_i\}$, without revealing the observables choices and the corresponding outcomes. Examples are given in Table \ref{TableS3}.
	
	\item[S4(a).] {\em Bob's check of Alice's honesty in S4:} Alice may cheat by including some pairs of runs in the list $\{R_i\}$, none of which satisfies either or both of the constraints $A_{i_1}=A_{i_2}$ and $a_{i_1}\neq a_{i_2}$. From the list $\{R_i\}$, Bob randomly selects some pairs, and asks Alice to announce for them her measurement bases $(A_{i_1},A_{i_2})$ as well as the corresponding outcomes $(a_{i_1},a_{i_2})$. By comparing with his own measurement results, for runs $(i_1,i_2)$, Bob can check whether Alice's list $\{R_i\}$ is genuine or not {\em i.e.}, whether all the pairs $(i_1,i_2)$ satisfy $A_{i_1}=A_{i_2}$ and $a_{i_1}\neq a_{i_2}$. If the list $\{R_i\}$ is genuine, due to the second and the third condition of (\ref{hardy2q}), Bob would never find `$U=+1$' (or, `$D=+1$') if Alice's result is $A_{i_1}/A_{i_2}=`D=+1'$ (or, `$U=+1$'). Also Bob can perform a measurement associated to the observable $A_i$ on the qubits he received from Alice for those declared pairs of runs from the list $\{R_i\}$, and compares the outcomes with $a_i$'s declared by Alice. In this way he can check whether the list $\{R_i\}$ announced by Alice is genuine or not. If the list $\{R_i\}$ is genuine, then his measurement outcomes must be equal to $a_i$ for each of those declared pairs of runs. For example, if instead of the pair of states $( \ket{u}  \ket{u},\ket{u^\perp}  \ket{u})$, Alice selects the pair $( \ket{u}  \ket{u},\ket{d^\perp}  \ket{u})$ (although upon Bob's query, Alice declared the former pair), then by performing measurement in the basis $ \ket{u},\ket{u^\perp}\}$ on Alice's qubit, Bob can detect the genuinity of the pair.
	
	\item[S5.] {\em Runs offered by Bob for OT scheme:} From remaining pairs of runs, Bob announces all the pairs of runs (say, $\{R'_i\}\subset \{R_i\}$), in which his measurement pairs are different {\em i.e.} $B_{i_1}\neq B_{i_2}$. Thus, in the ideal case, for the list $\{R'_i\}$, the possible pairs of joint states of Alice and Bob will be one of the last four pairs of (\ref{forOT}). Examples are given in Table \ref{TableS3}. 
	For remaining pairs $\{R_i\}\setminus \{R'_i\}$ of runs Bob declare his choice of measurement $B_{i_1}$ and $B_{i_2}$.  
	\item[S5(a).] {\em Alice's check of Bob's honesty in S5:} By performing proper measurement on Alice's qubits, Bob can learn Alice's choice of measurement bases with probability $p$ for those pairs of runs $\{R'_i\}$. Therefore, Bob may cheat by announcing only those $\{\widetilde{R'}_i\}$ ($\subset \{R'_i\}$) pairs of runs in which he knows the Alice's measurement bases after performing the measurement on Alice qubits. But this will cause a very low frequency for the list $\{\widetilde{R'}_i\}$ and Alice can easily check that. For example, in the optimal case ({\em i.e.}, for $q=q_{max}$) the list $\{R'_i\}$ contains $50\%$ of the pairs from the list $\{R_i\}$ whereas, the list $\{\widetilde{R'}_i\}$ contains only $30.9\%$ of the pairs of $\{R_i\}$. Also, for the pairs $\{R_i\}\setminus \{R'_i\}$ runs Alice would never find `$U=+1$' (or, `$D=+1$') if Bob's measurement is $B_{i_1}/B_{i_2}=D$ (or, $U$). This is due to the second and the third conditions of (\ref{hardy2q}).
	
	Bob can cheat in step {S{\bf 5}}
	by including some pairs of runs in the list $\{R'_i\}$ in which his observables pairs do not satisfy the condition $B_{i_1}\neq B_{i_2}$. From the list $\{R'_i\}$, Alice randomly selects some pairs, and asks Bob to announce for them his measurement bases $(B_{i_1},B_{i_2})$. 
	By comparing with her own measurement results, for runs $(i_1,i_2)$, Alice can check whether Bob's list $\{R'_i\}$ are genuine or not. If it is genuine, due to the second and the third conditions of (\ref{hardy2q}), again Alice would never find `$U=+1$' (or, `$D=+1$') if Bob's measurement is $B_{i_1}/B_{i_2}=D$ (or, $U$). Thus, for example, Alice can detect the malicious pair $( \ket{u}  \ket{u},\ket{u^\perp}  \ket{u})$ (or, $( \ket{d}  \ket{d},\ket{d^\perp}  \ket{d})$) according to the aforesaid property.
	
	\item[S6.] {\em OT announcement by Alice:} From remaining pairs of runs $\{R'_i\}$, Alice selects one pair of runs, say $R_{i^*}=(i^*_1,i^*_2)$, according to the message she wants to convey to Bob (say, ``Alice chooses $(A_{i^*_1},A_{i^*_2})=(U,U)$ for communicating the bit value $0$, while she chooses $(A_{i^*_1},A_{i^*_2})=(D,D)$ for communicating the bit value $1$"), and announces only the index $i^*$ for OT. Thus, for the announced index $i^*$, in the ideal situation, the possible pairs of joint states of Alice and Bob will be one of the pairs from last four pairs of (\ref{forOT}).
	
	\item[S7.] {\em Message recovery by Bob:} Bob performs measurements in $X_{j}=B_{i^*_j}$ bases 
	on the $i^*_j$-th qubit (for $j=1,2$), he received from Alice in step S2. Thus, because of properties of the list $L^+$ the settings $(X_1,X_2)=(B_{i^*_1},B_{i^*_2})$ can be one of the two pairs $(U,D)$, or $(D,U)$ (this is due to the condition $B_{i^*_1}\neq B_{i^*_2}$ given in S5) - as appeared in the four possible pairs of states of Alice and Bob after the OT announcement by Alice. Note that when the pair of joint states is $( \ket{u}  \ket{u},\ket{u^\perp}  \ket{d})$ then, in the measurement of $X_1=U$ and $X_2=D$ on the two states of Alice will respectively give rise to the output states $ \ket{u}$ `with probability $1$' and $\ket{d}(\ket{d^\perp})$ `with probability $|\beta|^2(|\alpha|^2)$'. But, if the pair of joint states is $(\ket{d^\perp}  \ket{u},\ket{d} \ket{d})$ then, in the measurement $(X_1,X_2)=(U,D)$ on the two states of Alice will respectively give rise to the output states $\ket{u} (\ket{u^\perp})$ `with probability $|\beta|^2(|\alpha|^2)$' and $\ket{d}$ `with probability $1$'. Here one should note that the result $(+1 +1)$ give him (Bob) no clue about the Alice's measurement choice. But the result $(+1,-1)$ indicates the first case ({\em i.e.}, Alice's measurement basis is $U$) and the result $(-1 , +1)$ the second one ({\em i.e.}, Alice's measurement basis is $D$), with certainty. Similar will be the case with the rest two pairs $(\ket{u^\perp} \ket{d}, \ket{u}  \ket{u})$ and $( \ket{d}  \ket{d},\ket{d^\perp}  \ket{u})$, where Bob performs measurements $(X_1,X_2)=(D,U)$ on Alice's pair of qubits $(i^*_1,i^*_2)$. Thus, in general the measurement $X_j$ on the corresponding Alice's state $i^*_j$ can give a $-1$ outcome, only if $X_j=A^c_{i^*_j}$, where $A^c_{i^*_j}=U$ if $A_{i^*_j}=D$ and $A^c_{i^*_j}=D$ if $A_{i^*_j}=U$. The probability $P_B(-1)$ of getting at least one $-1$ outcome in the measurement of $(X_1,X_2)$ on the respective Alice's pair of states $(i^*_1,i^*_2)$ is equal to $|\alpha|^2(=p)$. Therefore, with probability $p$ Bob can learn Alice's basis choice, whereas Alice has no information whether Bob succeeds or not.
\enu
Thus, by now we have succeeded in establishing a secure $p$-OT scheme between Alice and Bob.

\section{Realistic noise robustness of the scheme}
Above prescription of the $p$-OT problem refers to an unfeasible, noise-free realization of the Hardy's test. But, in reality it is quite evident that the last three joint probabilities of (\ref{hardy2q}) will never be equal exactly to zero due to the imperfectness of the quantum resources used. Thus, it is worthwhile to investigate the permissible range of noise for a faithful realization of the scheme. Here we adopted the method describe in Ref. \cite{RWZ15} for noise robustness analysis. In presence of noise Hardy's conditions (\ref{hardy2q}) turn out as
\be\label{hardy2qn}
\begin{split}
	P(+1,+1|U,U) &>q-3\epsilon>0,\\
	P(+1,+1|U,D) &\leq \epsilon,\\
	P(+1,+1|D,U) &\leq \epsilon,\\
	P(-1,-1|D,D) &\leq \epsilon.	
\end{split}
\ee
for some small $\epsilon>0$. It was shown in Ref. \cite{RZS12} that for $q = q_{max}$ the Hardy
experiment (\ref{hardy2qn}) can provide a DI self test of the associated Hardy's correlation for $0\leq \epsilon \lesssim 0.2$. Hence, our $p$-OT scheme based on Hardy's correlation is also device-independent for $q = q_{max}$ and it is robust against noise for $0\leq \epsilon \lesssim 0.2$. In order to illustrate what orders of perfectness can be observed in real experiments, we refer to the setup of each subsystem with Hardy's measurements and assume
that the entangled state sources admixture with the white noise and produce Werner like states,
\be\label{werner}
\rho(\eta)=\eta \proj{\psi^H}+\frac{1-\eta}{4}\mathbb{I}_2\otimes \mathbb{I}_2.
\ee

Note that, the presence of noise is barring the users to conduct a perfect Hardy experiment (\ref{hardy2q}). It is therefore more viable to test the violation of the following LR-inequality derived from Hardy's conditions (\ref{hardy2q}) 
\be
	\label{hardyi}
P(+1,+1|U,U) - P(+1,+1|U,D)-P(+1,+1|D,U)-P(-1,-1|D,D)\leq 0.
\ee
This is nothing but, rearrangement of the expression of the well known CH inequality \cite{CH74}. The noisy correlation (\ref{werner}) reduces the Hardy experiment (\ref{hardy2q}) related four joint probabilities to be
\be	\label{Prob1} \begin{split}
	P(+1,+1|U,U)&=\eta q + \frac{1-\eta}{4} ,	\\
	P(+1,+1|U,D)&=\frac{1-\eta}{4},\\
	P(+1,+1|D,U)&=\frac{1-\eta}{4},\\
	P(-1,-1|D,D)&=\frac{1-\eta}{4},
	\end{split} \ee
where $\displaystyle q=\dfrac{|\alpha|^4(1-|\alpha|^2)}{1+ |\alpha|^2}$, $0<|\alpha|^2<1$.
The above set of probabilities provides a violation of the inequality (\ref{hardyi}) for \be\label{noise}
\eta > \dfrac{1}{1+2q}.
\ee
In particular, for $q=q_{max}$, $\eta>0.847214$ for a desired violation of the inequality (\ref{hardyi}). If we demand DI self test for the Hardy correlation $\ket{\psi^H}$ we must have $\dfrac{1-\eta}{4}<\epsilon \lesssim 0.2$ i.e., $\eta>0.92$. Thus, the description works, device independently, if the visibility of the shared Hardy states is above $92$\%. 

\begin{center}
	\begin{figure}
		\includegraphics[width=0.5\linewidth]{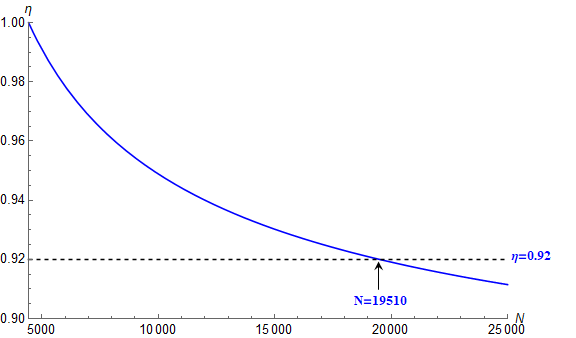}
		\caption{Plot between $\eta$ and $N$ for $q=q_{max}$ with $99.97\%$ faithfulness.}
		\label{fig1}
	\end{figure}
\end{center}

In practical, the users can not conduct the experiment with infinitely many number of runs. Thus, instead of probabilities, they learn nothing but the relative frequencies of the events based on number (finite) of runs they have conducted in the experiment and hence introduced some uncertainties. Therefore, it is justified to ask, what is the range of $\eta$, for which the $p$-OT scheme functions with reliability, say, 99.7\% for $N$ number of runs of the experiment. We assume that the probability density function $\rho(f|P)$ that the relative frequency $f$ found in a real experiment is equal to probability $P$ is Gaussian i.e., $\displaystyle \rho(f|P) =\dfrac{1}{\sqrt{\left( 2\pi \sigma^2\right)}}e^{-\dfrac{(f-P)^2}{2\sigma^2}}$, with standard deviation (SD) $\displaystyle \sigma \approx N^{-\frac{1}{2}}$. This is a reasonable assumption for the distributions which are not very far from their respective means and for sufficiently large number of runs $N$. To achieve the desired faithfulness of the scheme, the LHS of the inequality (\ref{hardyi}) should to be
larger than $3\times 2\sigma$. Here, factor $3$ stands for the desired faithfulness of the scheme and $2=\sqrt{4}$ since the inequality (\ref{hardyi}) contains four joint probabilities. For strong violation of the inequality (\ref{hardyi}) with desired reliability, we must have  
\be
\begin{split}
\eta q+ \frac{1-\eta}{2}  &> 6\sigma \\
\text{or,~~}\eta &>\frac{1}{2q+1}\left( 1+\frac{12}{\sqrt{N}}\right),
\end{split}
\ee
which demands $N>4428$ for $q=q_{max}$. If the visibility of the produced Hardy states is  $95$\% i.e., $\eta=0.95$ then we required $N>9783$ runs of the experiment for desired faithfulness. A comparative study between $\eta$ and $N$ for $q=q_{max}$ is shown in the Figure \ref{fig1}. 

\section{Conclusions} The ideal Hardy's test (\ref{hardy2q}) for maximum probability of success $q_{max}=\frac{5\sqrt{5}-11}{2}$ {\em i.e.}, for $|\alpha|^2=\frac{\sqrt{5}-1}{2}$, is fully device independent(DI) \cite{RZS12}, hence the present quantum scheme of $p$-OT is also DI for $p=\frac{\sqrt{5}-1}{2}$. As $\frac{\sqrt{5}-1}{2}\approx 0.618034 \in (0, 1)$, so one can construct a secure protocol for Rabin's OT {\it i.e.}, $\dfrac{1}{2}$-OT by using our $p$-OT protocol for $p=\frac{\sqrt{5}-1}{2}$, and security of which is evidently DI. By providing a secure quantum scheme for $p$-OT, we have thus answered a long standing open problem \cite{Lo97,He12} that, other than the quantum key distribution whether there is any quantum cryptographic protocols which is unconditionally secure.

As the protocol is essentially based on ``low frequencies of some lists", which is not guaranteed in stochastic situations, its security is only highly probable, and perhaps in the limit of infinitely many runs fully secure. Thus, the security of our scheme achieved its unconditional goal asymptotically. 

\section{Acknowledgments}

We thank Sibasish Ghosh and Marek \.{Z}ukowski for stimulating discussions and helpful comments for preparing the manuscript.

 \end{document}